\documentclass[10pt,a4paper]{book}
\usepackage{perugia}
\begin{document}

\talktitle{The AGILE archive and data processing at the ASI Science Data Center}

\talkauthors{Elisabetta Memola \structure{a}, 
             Paolo Giommi \structure{a}}

\authorstucture[a]{ASI Science Data Center,
                   Via Galileo Galilei~1, 
                   00044~Frascati, Italy}

\shorttitle{AGILE at the ASI Science Data Center} 

\firstauthor{E. Memola, P. Giommi}

\def\topfraction{.99}
\setcounter{bottomnumber}{1}
\def\bottomfraction{.3}
\setcounter{totalnumber}{3}
\def\textfraction{.01}
\def\floatpagefraction{.9}
\setcounter{dbltopnumber}{2}
\def\dbltopfraction{.7}
\def\dblfloatpagefraction{.5}

\begin{abstract}

AGILE (Astro-rivelatore Gamma ad Immagini LEggero) is a Small Scientific Mission of
the Italian Space Agency (ASI) with a Science Program open to the national and international
community. Its main goal is to develop and operate a scientific satellite devoted to Gamma-ray
(30\,MeV$-$50\,GeV) and hard X-ray (10$-$40\,keV) Astrophysics during the years 2005$-$2007.
ASI plans to handle AGILE data through the ASI Science Data Center 
in collaboration with the AGILE Team.

\end{abstract}

\section{AGILE and the ASI Science Data Center} 

AGILE, the 4$^{th}$ Gamma-ray satellite, is a bridge between the Compton Gamma Ray Observatory (CGRO),
switched off in 2000, and the Gamma Ray Large Area Space Telescope (GLAST) to be launched at the end
of 2006. 

The AGILE instrument is designed to detect and image Gamma-ray and hard X-ray photons
by means of the Gamma Ray Imaging Detector (GRID) and the hard X-ray imager Super-AGILE (SA).

GRID, sensitive in the energy range 30\,MeV$-$50\,GeV, is made
of 12 Si-W planes and the Mini Calorimeter, sensitive in the energy range 0.3$-$100\,MeV and
positioned at the bottom of the instrument. 

SA, sensitive in the energy range 10$-$40\,keV, with its 4
Si-detectors and the ultra-light coded mask system is positioned on top of the first GRID tray.

The ASI Science Data Center (ASDC) will host the 
AGILE Data Center (ADC). The ADC includes the AGILE Team Processing
Group and the AGILE Science Support Group. AGILE data from the Malindi Ground Station will be
received by the Operational and Control Center (AOCC) in Italy and then 
they will be transfered to ADC. \\

\noindent
The ADC@ASDC will be in charge of the following tasks \cite{plan}:

\begin{itemize}
\item
Running the Quick Look Analysis
\item
Running the Standard data reduction Analysis
\item
Performing, when necessary, the Interactive data Analysis
\item 
Managing Announcement of Opportunities
\item
Contributing to the AGILE management of the Pointing Program
\item
Archiving the data (raw, cleaned and calibrated, scientific)
\item
Distributing the data to the scientific community 
\item
Providing scientific support to the users community
\item
Interfacing the project for both data and proposals \\ via dedicated web pages 
\item
Providing the standard software support for data analysis
\end{itemize}

\begin{figure}[h]
\vspace{18.3cm}
\includegraphics{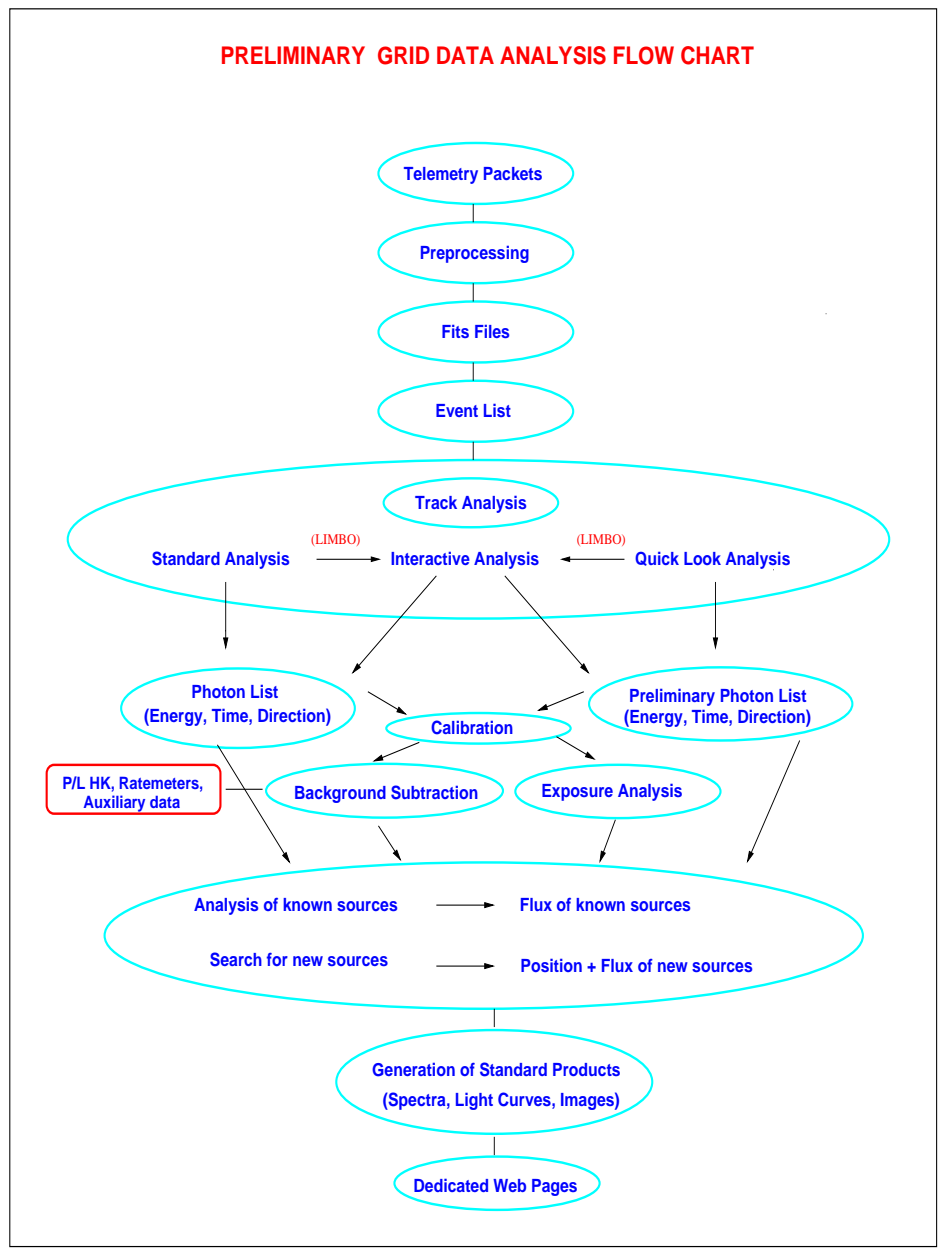}
\vspace*{-1.3cm}
\caption
{Preliminary flow chart describing the pipeline that will drive the automated data reduction
and analysis system for GRID data at ASDC.}
\label{memola1}
\end{figure}

\noindent
The pipeline that will be running at ASDC will drive the automated data analy\-sis for both
GRID and SA data. 

Figure~\ref{memola1} shows a preliminary flow chart concerning the GRID 
data analysis. The software modules needed to process the data are under development
by the AGILE Team, while the pipeline itself will be built by professional programmers
of the AGILE Science Support Group at ASDC, in collaboration with the AGILE Team.
The GRID telemetry packets will be pre-processed in order to get files in the standard
fits format. The event list undergoes a complex process of background subtraction and track
reconstruction analysis in order to get a cleaned $\gamma$-photon list. 
Then, standard products as spectra, light curves and images will be made available to the
scientific community via a set of dedicated web pages.
AGILE data will be part of the permanent multi-mission interactive archive at ASDC.

\section{AGILE and blazars} 
\label{seds}

The scientific goals of the AGILE mission include the detailed study of
Active Galactic Nuclei. 

\begin{figure}[t]
\vspace{10.0cm}
\includegraphics{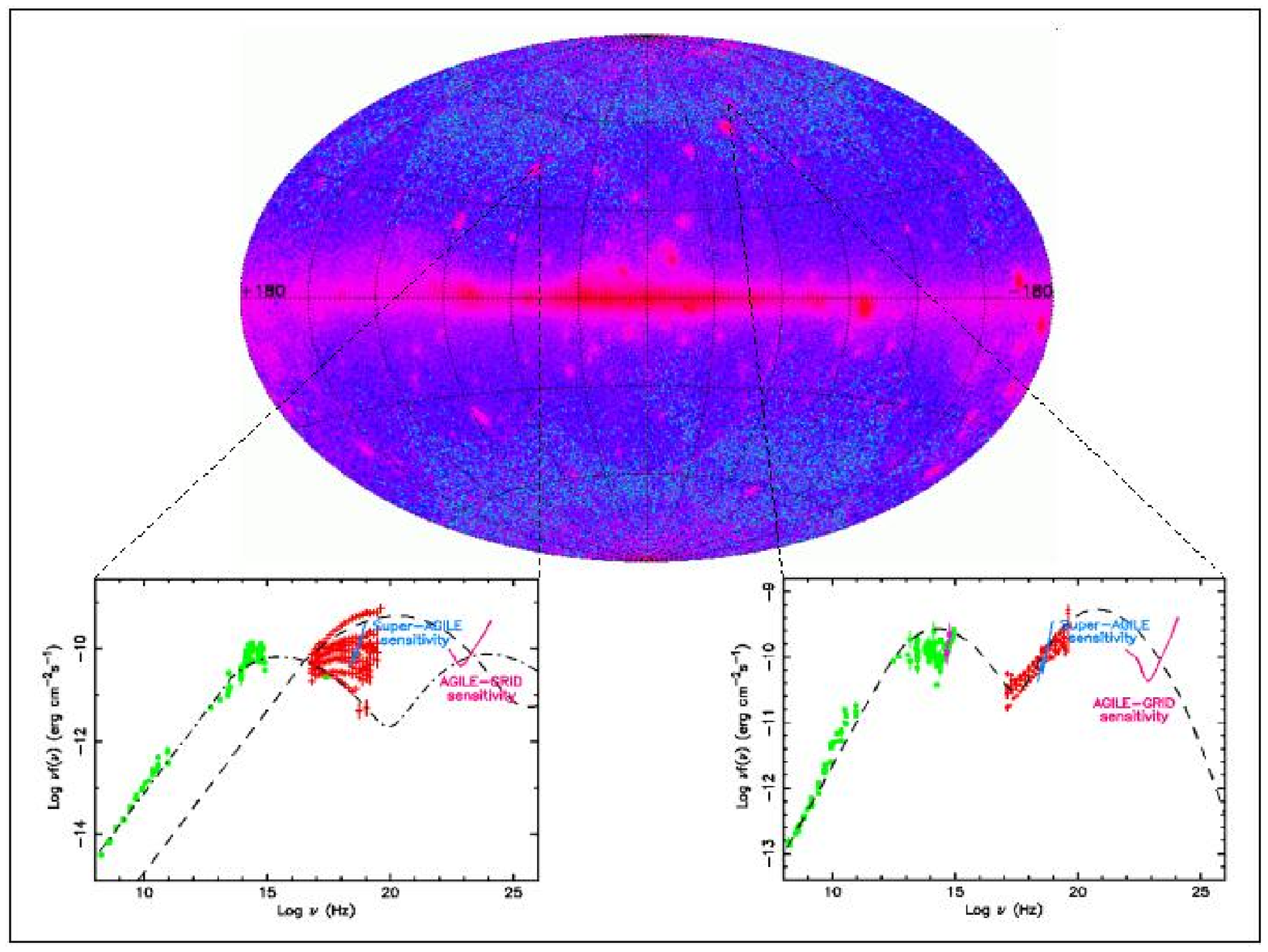}
\vspace*{-0.2cm}
\caption
{{\em Top}: AGILE simulated all-sky intensity map above 100\,MeV assuming the complete
sky coverage with 6 pointings lasting 4 weeks each \cite{agile}.
{\em Bottom}: Spectral Energy Distributions of the blazars Mkn~501 ({\em left}) and 3C~273 ({\em right}) $-$ 
(for more details see \S~\ref{seds} and http://www.asdc.asi.it/blazars/).
}
\label{memola2}
\end{figure}

Figure~\ref{memola2} ({\em top}) shows
the simulated all-sky intensity map above 100\,MeV \cite{agile}
as well as the Spectral Energy Distributions (SEDs) of 
the blazars Mkn~501 ({\em bottom, left}) and 3C~273 ({\em bottom, right}). 
The multi-frequency data have been retrieved
from the NASA/IPAC Extragalactic Database (NED), 
the second version of the Guide Star Catalog (GSC-II) and from {\em Beppo}SAX.
By accounting for AGILE-GRID and
Super-AGILE limiting sensitivity we plan to select the blazars
currently known that might be suitable candidates for AGILE observations.
According to the predictions of a Synchrotron-Self Compton (SSC) spectral model
({\em dashed}, {\em dashed-dotted lines}),
Mkn~501 and 3C~273 would be visible to both AGILE-GRID and Super-AGILE.
The SEDs of all the blazars observed by {\em Beppo}SAX
during its first five years of operations 
\cite{sed}
and the corresponding SSC predictions together with the plotted 
AGILE-GRID and Super-AGILE limiting sensitivity can be found at 
the following web address: http://www.asdc.asi.it/blazars/

\end{document}